\documentclass[a4paper,usenatbib]{mn2e}
\usepackage{journals}
\usepackage{natbib}

\usepackage[T1]{fontenc}
\usepackage{ae,aecompl}

\newcommand{\SoS}{{Solar system }}

\newcommand{\EQ}[1] {equation~(\ref{#1})}
\newcommand{\SEC}[1] {Section~\ref{#1}}

\newcommand{\FIG}[1] {Figure~\ref{#1}}

\newcommand{\VEC}[1] {{\boldsymbol{{ #1}}}}
\newcommand{\MX}[1] {{\mathbfss{{#1}}}}
\newcommand{\D} {\mathrm{d}}

\usepackage{graphicx}	
\usepackage{amsmath}	
\usepackage{amssymb}	
\usepackage{color}

\title[Explore unknown mass using PTA]{A dynamical approach in exploring the unknown mass in the Solar system using pulsar timing arrays}

\author[Y. J. Guo et al.]{
Y. J. Guo$^{1}$, 
K. J. Lee$^{1,2}$\thanks{E-mail: kjlee@pku.edu.cn}
and R. N. Caballero$^{3}$
\\
$^{1}$Kavli Institute for Astronomy and Astrophysics, Peking University, Beijing 100871, People's Republic of China\\
$^{2}$National Astronomical Observatories, Chinese Academy of Sciences, Beijing 100012, 
China\\
$^{3}$Max-Planck-Institut f\"ur Radioastronomie, Auf Dem H\"ugel 69, Bonn 
53121, Deutschland}

\date{Accepted XXX. Received YYY; in original form ZZZ}

\pubyear{2017}

\begin{document}
\label{firstpage}
\pagerange{\pageref{firstpage}--\pageref{lastpage}}
\maketitle

\begin{abstract}
The error in the Solar system ephemeris will lead to dipolar
correlations in the residuals of pulsar timing array for widely separated 
pulsars.  In this paper, we utilize such correlated signals, and construct a 
Bayesian data-analysis framework to detect the unknown mass in the Solar system 
and to measure the orbital parameters. The algorithm is designed
to calculate the waveform of the induced 
pulsar-timing residuals due to the unmodelled objects following the Keplerian 
orbits in the Solar system. 
The algorithm incorporates a Bayesian-analysis suit
used to simultaneously analyse
the pulsar-timing data of multiple pulsars to search for coherent 
waveforms, evaluate the detection significance of unknown 
objects, and to measure their parameters. When the object is not detectable, our 
algorithm can be used to place upper limits on the mass.  
The algorithm is verified using simulated data sets, and cross-checked 
with analytical calculations. We also investigate the capability of future 
pulsar-timing-array experiments in detecting the unknown objects. We expect that 
the future pulsar timing data can limit the unknown massive objects in the Solar 
system to be lighter than $10^{-11}$ to $10^{-12}$ $M_{\rm \sun}$, or measure
the mass of Jovian system to fractional precision of $10^{-8}$ to $10^{-9}$.
\end{abstract}


\section{Introduction}

The clock-like rotational stability
of millisecond pulsars makes them the 
the most accurate celestial clocks known.
Through the process of pulsar timing \citep[e.g.][]{LK05,HEM06b},  
millisecond pulsars are powerful tools
for a wide range of scientific problems. In the process of pulsar
timing, the time of arrivals (TOAs) recorded at the observatory are
transferred to the pulsar frame. The differences between the
observed TOAs and the model predictions form the timing residuals. The
physical processes not modelled will leave their fingerprints in the timing
residuals. For processes affecting all the pulsars, they introduce the
correlated signals between widely separated pulsar pairs. Such spatial
correlations have profound applications. For example, one can detect
the gravitational waves \citep[GW;][]{HD83},
investigate the stability of reference terrestrial time standards \citep{HCM12}, 
and study the \SoS ephemeris~\citep{Champion10}. These applications
make use of so-called pulsar timing arrays (PTAs), which are 
an ensemble of pulsars, typically millisecond pulsars, in different sky positions \citep{FB90}.

The first step in converting the site TOAs to the pulse-emission time 
at the pulsar frame, is to refer
them to the \SoS barycentre (SSB) according to the SSB position with respect
to the Earth. In the common pulsar timing practice, the SSB position is
provided by the Solar system ephemeris~\citep{Standish98}. Errors in the
used ephemeris will then lead to inaccurate conversion of TOAs, and thus
induce the correlated timing residuals among all analysed pulsars. We can study 
the Solar system ephemeris by searching for such correlated residuals. 

\citet{Champion10} were the first to employ a PTA to constrain
the mass of planets in the \SoS. They fixed the orbits of the known planets using
the DE421 ephemeris \citep{Folkner09} and investigated the effects of perturbing the
input planetary masses on the TOAs. In this way,
they used the PTA data to constrain possible errors
in the Solar system ephemeris, and provide upper limits on the 
masses of planets (or planetary systems when satellites are in orbit).

Possible errors in the Solar system ephemeris may come from two aspects:
inaccurate mass or position of known objects, and the existence of
unmodelled objects (UMOs). \citet{Champion10} had studied the error in the
mass of known planets. In this paper, we want to explore the unknown
objects in the \SoS by pulsar timing, to detect their signal or put upper limits on their masses. 
The term UMOs here refers to any unknown objects revolving around the SSB,
such as dark matter clumps \citep{LZ05, PP13}, small asteroids \citep{ST16},
strangelets \citep{WXM07}, cosmic strings \citep{BOS14}, tiny primordial black holes or other
unidentified massive objects. The studies of timing residuals induced
by the ephemeris can help us understand the noise budget of PTAs, which
is of central importance for the GW detection with PTAs. Furthermore, a better Solar system
ephemeris may also improve the precision of deep space missions.

In the current paper, we model the UMOs with Keplerian orbits, calculate the 
induced timing residuals, and perform parameter estimation using
Bayesian inference. We describe our methods in \SEC{sec:methods}, and use 
simulations to verify our algorithm in \SEC{sec:sim}. We
analytically calculate the PTA sensitivity to the UMOs and investigate the 
capability of future PTA experiments in \SEC{sec:analytical}.
Discussions and conclusions are made in \SEC{sec:conclusion}.

\section{Methods}
\label{sec:methods}

We use a model-based Bayesian data-analysis method to measure the  
mass of UMOs using PTA data. There are two major components for the 
Bayesian inference, the signal model (in \SEC{sec:waveform}) and the likelihood 
model (in \SEC{sec:prob}). 

\subsection{Pulsar-timing residuals induced by UMOs}
\label{sec:waveform}

Pulsar timing residuals $\VEC{R}$ can be described as the 
sum of three sources: 
the signal $\VEC{\Delta R}$, 
induced by unmodelled processes such as the orbit of a UMO,
the signal $\VEC{s}$, induced by imperfectly 
modelled timing parameters, 
and the signal $\VEC{n}$, from noise processes. That is
\begin{equation}
	\VEC{R}=\VEC{\Delta R}+\VEC{s}+\VEC{n}\, .
	\label{eq:allres}
\end{equation}

In most of the pulsar-timing procedures, one uses the clock corrections published 
by the Bureau International 
des Poids et Mesures (BIPM)\footnote{http://www.bipm.org/}, Earth orientation parameters 
and Solar system ephemeris to correct the TOAs seen at the telescope site to the 
TOAs as seen in the pulsar rest frame. In this way, any objects not included in 
the ephemeris will introduce signals in the pulsar timing residuals. 
Such signals have a dipolar spatial 
correlation, different from the monopolar correlation induced by clock errors \citep{HCM12}.
In this paper, we focus on the leading-order effects of the UMOs under the following two 
assumptions.

{\bf (i)} We assume that the UMO follows a point-mass Keplerian orbit around the 
SSB. In particular, we focus on bound systems, of which orbits are elliptic. The 
major acceleration of the UMO is thus due to the Sun, and we neglect the 
higher-order effects, such as the perturbations from objects except the Sun in the 
Solar system, the post-Newtonian corrections, and tidal forces.

{\bf (ii)} The UMOs induce periodic motion of the SSB. 
Such motion will 
contribute to the pulsar-Earth distance and lead to an extra geometric time 
delay in the pulsar TOAs
(i.e. the R\o mer delay as explained in \citet{EHM06}). We have neglected the 
higher-order effects due to the SSB motion (e.g.  parallax, gravitational 
redshifts, and Shapiro delay), as done by \citet{Champion10}.

\citet{Champion10} showed that the perturbation of Jupiter mass simply changes the position of the SSB, as we modelled. 
It is unclear if such perturbative approach is also valid for the other planets, especially the ones with inner orbits. 
Investigating the long-term evolution of the Solar system dynamics with full modelling is beyond the scope of this article. 
However, the data length is limited to only a couple of tens of years, such that the first-order treatment, i.e. UMOs induce SSB shifts, is a valid approximation.

The R\o emer delay $\Delta R$ associated with the displacement $\VEC{d}_{\rm s} 
$ of the SSB is
\begin{equation}
	\Delta R=-\frac{\VEC{d}_{\rm s} \cdot \VEC{p}}{c}\,,
	\label{eq:delay}
\end{equation}
where $\VEC{p}$ is the unit vector pointing to the direction of the pulsar, and 
$c$ is the speed of light.
As we neglected the interaction between UMO and other \SoS objects other than 
the Sun,
the displacement of the SSB caused by the UMO with mass $m\ll M_{\rm \sun}$ and 
position vector $\VEC{r}$ relative to the original SSB is \begin{equation}
	\VEC{d}_{\rm s}=\frac{m}{M} \VEC{r},
	\label{eq:ds}
\end{equation}
where $M$ is the total mass of the \SoS and can be well approximated by the 
solar mass, such that $M\simeq M_{\sun}$.

The Keplerian orbit of a UMO is modelled with seven
parameters ($\xi_{i}, i=1\dots7$), which fully 
determine the induced pulsar-timing residuals. The $\xi$ parameters 
contain the mass of the UMO 
and six orbital elements. The orbital elements determine the geometry 
of orbit, and are the semi-major axis, $a$, eccentricity, $e$, longitude of the 
ascending node, $\Omega$, orbital inclination angle, $\iota$, argument of 
perihelion $\omega$, and phase at reference epoch, $\varphi_0$.

For elliptic orbits, the radial distance of a UMO to the centre of mass, $r$, can 
be expressed in terms of the eccentric anomaly, $u$, as 
\begin{equation}
	r=a(1-e\cos u)\,.
\end{equation}
The evolution of $u$ with time $t$ satisfies
\begin{equation}
u-e\sin u=2\pi ft-\varphi_{0} \,,
\end{equation}
where $f$ is the orbital frequency. The latter follows Kepler's third law, such that
\begin{equation}
f=\frac{1}{2\pi}\sqrt{\frac{GM}{a^{3}}}\,.
\end{equation}
Using the true anomaly $\nu$, the position vector $\VEC{r}_{0}$ of the UMO in the 
orbital plane becomes
\begin{equation}
	\VEC{r}_0=    \begin{pmatrix}
		r\cos \nu \\
		r\sin \nu \\
    0
		\end{pmatrix}\,,
\end{equation}
and the true anomaly is defined as
\begin{equation}
	\nu=2 
	\arctan\left[\sqrt{\frac{1+e}{1-e}}\tan\left(\frac{u}{2}\right)\right]\,.
	\label{eq:tranom}
\end{equation}
We transform the position vector to the ecliptic coordinate using rotation 
matrix as computed from the Euler angles of orbital elements, such that
\begin{eqnarray}
		\VEC{r} &=& {\MX R}_z(-\Omega) {\MX R}_x(-\iota) {\MX R}_z(-\omega) 
		\VEC{r}_0\,, \label{eq:rot}\\
		\MX{R}_{\rm z}(-\Omega)  &=&  \begin{pmatrix}
		\cos \Omega & -\sin \Omega & 0 \\
    \sin \Omega & \cos \Omega & 0 \\
    0 & 0 & 1
		\end{pmatrix},\\
		\MX{R}_{\rm x}(-\iota)  &=& \begin{pmatrix}
		1 & 0 & 0 \\
		0 & \cos \iota & -\sin \iota \\
		0 & \sin \iota & \cos \iota
		\end{pmatrix}.
\end{eqnarray}

The dependence of the timing residuals, $s$, on the timing parameters, 
$\VEC{\lambda}$, is usually non-linear, but the timing model can be linearized 
around the reference timing parameters, $\VEC{\lambda}_0$, to compute the timing 
residuals \citep{MT77, LK05, EHM06,vHL09}, which leads to
\begin{equation}
	s_i=\sum_{k}D_{ik}(\lambda_{k}-\lambda_{0, k})= \sum_{k}D_{ik} \delta 
	\lambda_{k}\,.
\end{equation}
Here, $i$ is the index of each epoch of observation, 
and $k$ is an index for the 
timing parameter.
$\VEC{D}$ is the design matrix (the coefficients of linearization), and 
$\delta\VEC{\lambda}$ is the deviation of timing parameters from the reference 
values. 

Unlike the deterministic signal $\Delta R$ and $s$, the noise components, $n$, 
in the timing residuals are random. We model them through the likelihood function 
as described in the next section. 

\subsection{The likelihood function and Bayesian inference}
\label{sec:prob}

We perform the parameter estimation using Bayesian inference. The Bayesian 
techniques had been studied extensively in the field of pulsar timing 
\citep{vHL09, LBJ14, Caballero16, LSC16}. Bayesian inference relies on 
converting the `probability of data' to the `probability of parameters' using
Bayes' theorem, 
\begin{equation} P(\VEC{
\zeta|X})=\frac{P({\VEC \zeta})P(\VEC {X|\zeta})}{P({\VEC X})}\,,
\end{equation}
where $\VEC X$ is the data, and $\VEC \zeta$ are the parameters to be inferred.  
$P(\VEC {X|\zeta})\equiv\Lambda$ is the likelihood function, i.e. the 
probability density function for the data 
given the parameters.
$P(\VEC {\zeta|X})$ is the posterior probability distribution, i.e. the probability 
density function for the parameters given the data set. The Bayesian evidence 
$P(\VEC X)$ is a normalization coefficient, defined as
\begin{equation}
	P(\VEC X)=\int P({\VEC \zeta})P(\VEC {X|\zeta}) d\VEC{\zeta}\,.
	\label{eq:evid}
\end{equation}
The prior probability distribution $P(\VEC \zeta)$ describes {\it 
a priori} belief about the distribution of the model's parameters. 

In the current paper, we assume the random noise in pulsar-timing residual of individual pulsars is a 
zero-mean Gaussian process. This approach had been taken by many authors. We 
refer the interested readers to \citet{LBJ14} or \citet{Caballero16} for the 
details of single-pulsar noise modelling. Here we only briefly outline the definitions.

Under the Gaussian assumption the noise components $\VEC{n}$ can be fully characterized using the covariance 
matrix, $\MX{C}$,
\begin{equation}
	P(\VEC{n|\Theta})=\frac{1}{\sqrt{(2\pi)^N|\MX{C}|}} \exp \left[ -\frac {1}{2} 
	\VEC{n}^{\MX T} \MX{C}^{-1} \VEC{n} \right]\,,
\end{equation}
where $N$ is the number of data points, $\VEC{\Theta}$ refers to the 
noise model parameters, and symbols $||$, $^{-1}$, and $^{\MX T}$ are the 
determinant, inversion, and transpose operation for matrices, respectively.

The noise processes in pulsar timing, are usually classified into three major 
parts, white-noise, red-noise, and the frequency-dependent-noise processes
\citep[see][for a review]{C13}. 
In this paper, we focus on the first two, the white noise and red noise. The noise power is 
additive, if the white noise and red noise are uncorrelated.  The noise 
covariance matrix becomes $\MX{C}=\MX{C}_{\rm w}+\MX{C}_{\rm r}$. 

The white noise is modelled as the TOA uncertainty $\sigma_i$ scaled by a 
systematic
factor (`Efac') , with $\sigma_{i}$ determined by the template fitting of pulse 
profile \citep{HVM04}.
We also include a possible independent source of white noise (such as jitter) which is modelled by Equad.
The white noise covariance matrix becomes
\begin{equation}
C_{{\rm w},ij}=
\begin{cases}
{\rm Efac}^2 \sigma_{i}^2+{\rm Equad}^{2}, &\text{if } i=j; \\
0, &\text{if } i\neq j.
\end{cases}
\end{equation}

The red noise is modelled as a wide-sense stationary Gaussian stochastic signal with a power-law 
spectrum~\citep{LBJ14},
\begin{equation}
 S_{\rm r}(f) =
 \begin{cases}
 \frac{A_{\rm r}^2}{f} \left(\frac{f}{f_{\rm c}} \right)^{2\alpha_{\rm r}}, &\text{if } f>1/T;\\
 0, &\text{if } f<1/T.
 \end{cases}
 \end{equation}
where $f_{\rm c}=1\, {\rm yr}^{-1}$ is a reference frequency,
and $T$ is the data length. 
The Fourier transform of the spectral density gives the temporal correlation of 
the red noise,
\begin{equation}
 C_{{\rm r},ij} =\int_{1/T}^\infty S_{\rm r}(f) \cos(2\pi ft_{ij}) {\rm d}f,
\end{equation}
where $t_{ij}$ is the time difference between the $i$-th and $j$-th epoch.

With all the ingredients, the likelihood function for the UMO problem is
\begin{eqnarray}
	\Lambda&\equiv & P(\VEC{R}|\VEC{\xi}, \VEC{\delta\lambda}, \VEC{\Theta}) 
	\nonumber \\
	&=&\frac{1}{\sqrt{(2\pi)^N|\MX{C}|}} e^{ -\frac {1}{2} (\VEC{R}-\VEC{\Delta 
	R}-\MX{D}\delta\VEC{\lambda})^{\MX T} \MX{C}^{-1} (\VEC{R}-\VEC{\Delta 
	R}-\MX{D}\delta\VEC{\lambda}) }\,.
	\label{eq:fullik}
\end{eqnarray}
The parameters that we are interested in are the orbital elements $\VEC{\xi}$.  
We can marginalise the likelihood function over the other parameters, which are referred to as ``nuisance parameters''.
The timing model parameters 
$\delta \VEC{\lambda}$ are linear, and they can be marginalised 
analytically~\citep{vHL09} giving
\begin{eqnarray}
	P(\VEC{R}|\VEC{\xi},\VEC{\Theta}) &=& \sqrt{\frac{|\MX{C}'^{-1}|(2\pi)^{M}}{|\MX{C}|(2\pi)^{N}}} \exp \left[ -\frac { \chi'^2 }{2} \right],  \label{eq:likli} 
\end{eqnarray}
with
\begin{eqnarray}
\MX{C}'&=&{\MX D}^{\MX T} \MX{C}^{-1}\MX{D} \,, \\
\chi'^{2}&=&(\VEC{R}-\VEC{\Delta R})^{\MX 
T}(\MX{C}^{-1}-\MX{G})(\VEC{R}-\VEC{\Delta R})\, , \\
\MX{G}&=&\MX{C}^{-1}\MX{DC}'^{-1}\MX{D}^{\MX T}\MX{C}^{-1}\,,
\end{eqnarray}
where M is the number of timing model parameters.

The noise parameters in $\VEC{\Theta}$ are nonlinear, and their marginalisation can only 
be performed numerically. We marginalise them in the stage of stochastic 
sampling of
the posterior. In this way, we fit the orbital elements and noise parameters 
simultaneously, with the parameters of timing model marginalised analytically.

Besides the likelihood function, the Bayesian analysis also depends on the prior 
distribution $P(\VEC{\Theta})$. For the parameter estimation, one should use the 
least informative prior, the Jeffreys prior \citep{Gregory2005}. The prior 
distributions are uniform for dimensionless parameters, while uniform in the 
log-space for the parameter with dimension. 
However, the logarithmic prior introduces an infinite-volume parameter space close to the origins, 
which makes the confidence level of upper limits invalid.  
As a consequence, the use of 
uniform priors is required to place reasonable upper limits \citep{Caballero16,LSC16}. 

In the following sections, we demonstrate our method by analysing simulated data 
sets. The sampling of posterior is carried out using the nested-sampling
Monte Carlo algorithm  \textsc{multinest} \citep{Feroz2009a}.
A paper where the presented method is applied to 
the first \emph{International Pulsar 
Timing Array} data \citep{VLH16},
is now in preparation.

\section{Demonstration and verification}
\label{sec:sim}

We simulate timing data for five pulsars from the IPTA
pulsar list \citep{VLH16}, namely PSRs J0437$-$4715, J1012$+$5307, 
J1713$+$0747, J1909$-$3744 and J2145$-$0750.
These pulsars have the lowest level of timing residuals and cover a wide 
distribution on the sky, so are suitable for our algorithm verification.
The parameters of timing noise injected in the simulated data are listed in 
Table~\ref{tab:sim}, where
the timing precisions are from the root-mean-square values of residuals in 
\citet{VLH16}.
The properties of the red noise are consistent with the values in
\citet{Caballero16} and \citet{LSC16}. 
We simulate the data with a cadence of 
two weeks and total length of ten years.
\begin{table}
	\centering
	\caption{The pulsars used in the simulation and the values of the
	injected noise parameters (timing precision $\sigma$, amplitude $A_{\rm r}$ and spectral index $\alpha_{\rm r}$ of the red noise).}
	\label{tab:sim}
	\begin{tabular}{cccc} 
		\hline
		\hline
		PSR & $\sigma (\rm \mu s$) & $A_{\rm r} (\mu s)$ & $\alpha_{\rm r}$ \\
		\hline
		J0437$-$4715 & 0.3 & 0.08 & -0.41\\ 
		J1012$+$5307 & 1.7 & 0.20 & -0.21\\ 
		J1713$+$0747 & 0.3 & 0.09 & -0.40\\ 
		J1909$-$3744 & 0.2 & 0.02& -0.11\\ 
		J2145$-$0750 & 1.2 & 0.31& -0.02\\ 
		\hline
	\end{tabular}
\end{table}

We address two scenarios here,

\textsc{case 1} If there are no detectable UMOs, we derive upper limits on their 
masses.

\textsc{case 2} If the UMO signal is strong, we measure their orbital elements.

We use the following recipe to simulate our data. \begin{enumerate}
\item Simulate the perfect TOAs for each pulsar using \textsc{tempo2}.

\item Add the white and red noise according to the noise parameters, where the 
	red noise is synthesised from the given spectrum using the fine frequency grid 
	of $4\times10^{-2}\, {\rm yr}^{-1}$.

\item For \textsc{case 2}, we inject the UMO induced signal.
\end{enumerate}

\begin{figure}
	\includegraphics[width=3.0in]{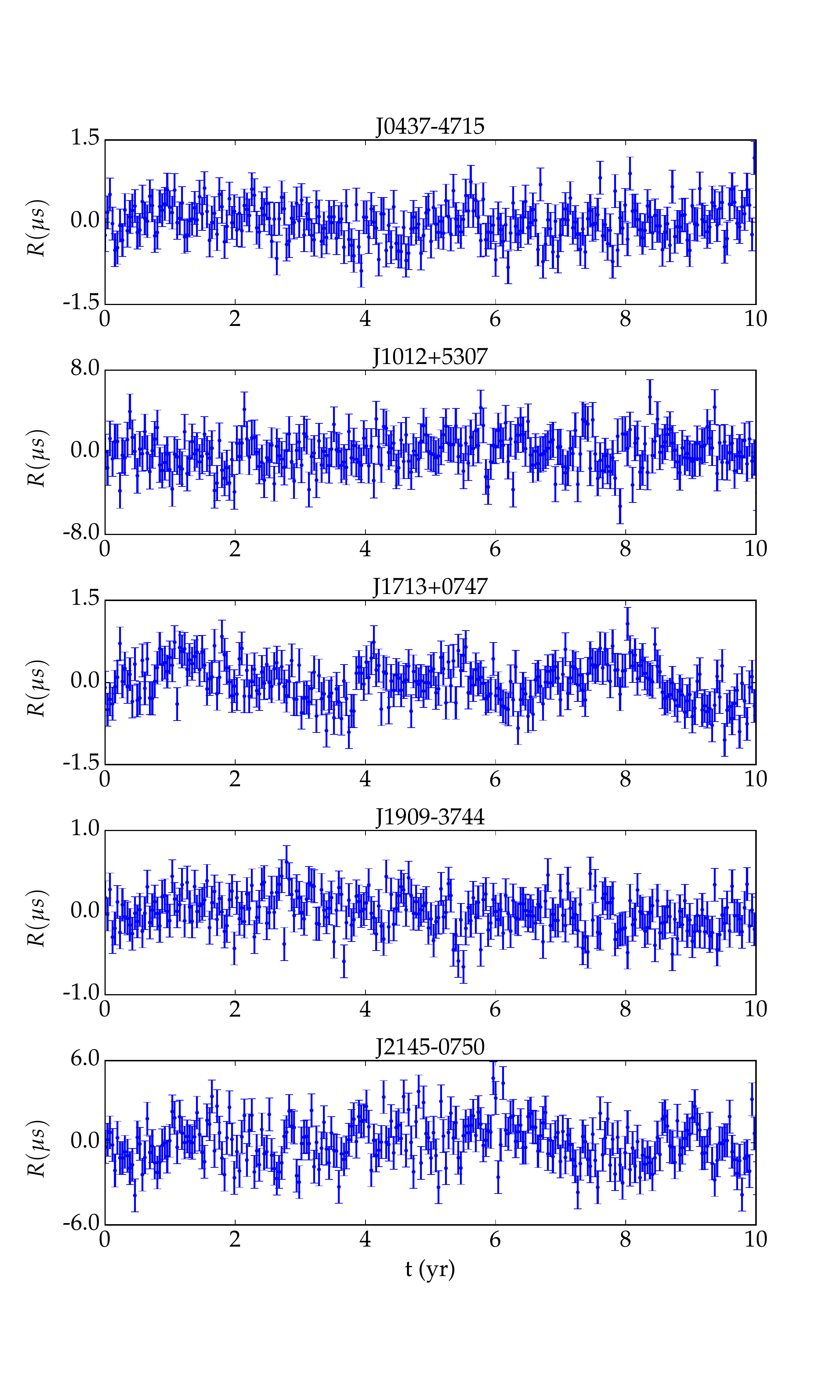}
	\caption{The simulated data in \textsc{Case 1}. We do not 
	inject the UMO signal in the data, and the structured waveform is due to the 
	red noise of each pulsar.}
    \label{fig:sim_data1}
\end{figure}

{\bf Data analysis and results of {\textsc case 1}}.
In this case, the UMO signal is not injected. We use Bayesian techniques to 
derive the upper limit for the mass of UMO. The timing residuals of the 
simulated data set are plotted in \FIG{fig:sim_data1}.

Our data analysis consists of two major steps. In the first step, we quantify if 
we \emph{detect} the UMO, while in the second step, we perform parameter 
\emph{inference}. We use Bayes factor ($K$) to evaluate the detection 
significance.  To do so, we need two Bayesian samplings, one using the model
including only the noise parameters and the other one using
the model including both noise parameters and UMO parameters. Then the Bayes 
factor $K$ is the ratio between the Bayesian evidence of the two model.  For the 
data in \FIG{fig:sim_data1}, we get $2\log K=0.2$.
Based on the interpretation of the Bayes factor by \citet{KR95},
this is an indication that the simpler model without the presence
of UMO signals is preferred, and so this
allows us with confidence to assume non-detection
for this data set and proceed to the next step.

\begin{figure*}
	\includegraphics[width=2\columnwidth]{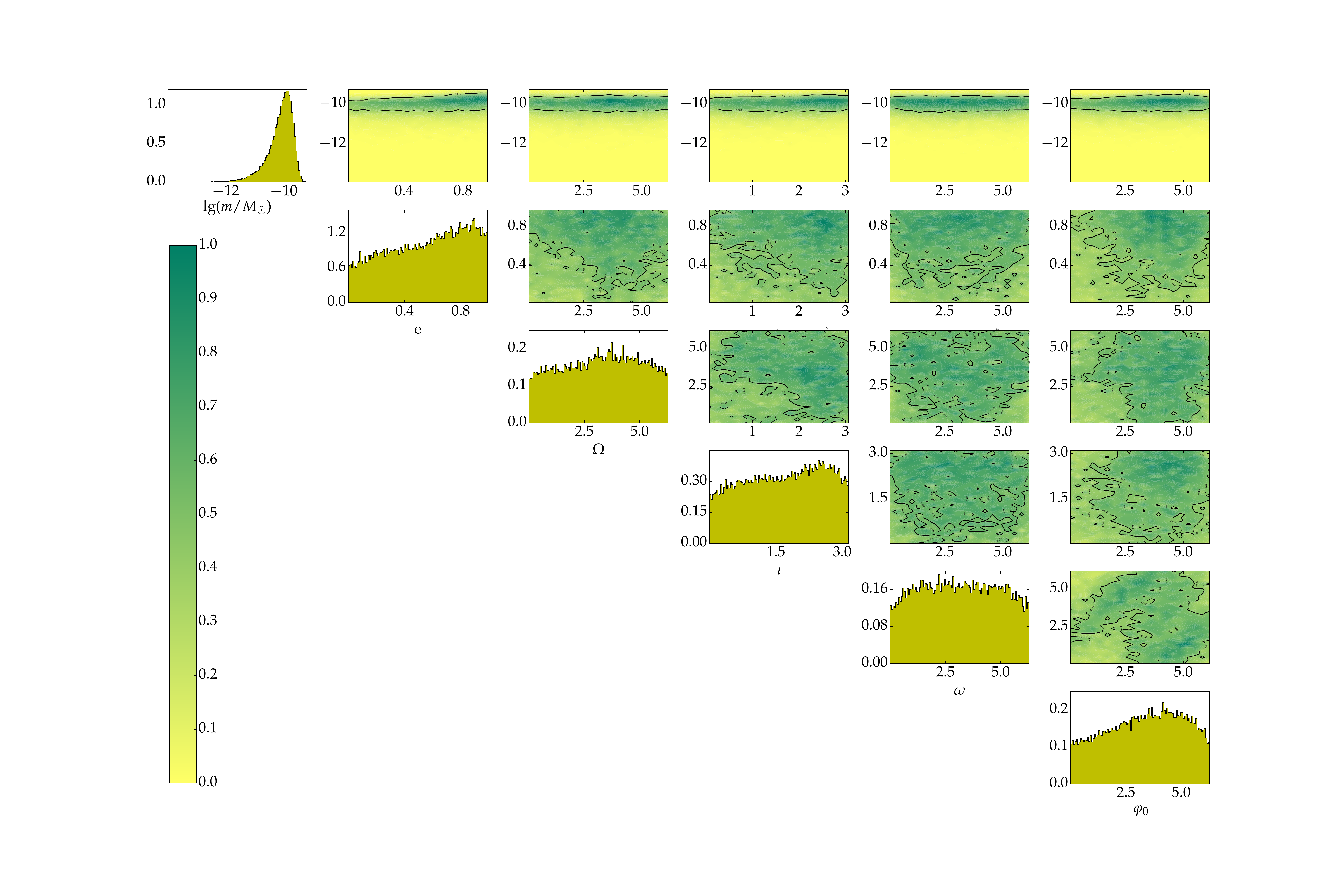}
		\caption{
		The marginalised posterior distribution of mass $m$ and orbital elements $e, \Omega, \iota, \omega, \varphi_{0}$ for the upper-limit analysis at $a \simeq 0.4\, $AU of the data set in \textsc{case 1}.  
The diagonal histogram plots are the 1D marginalised posterior distribution of the six parameters.
The upper triangular plots show the 2D marginalised posterior, which represent the correlations between parameter pairs.
The colour scale indicates the probability density that is rescaled to make the maximum value equal to 1,
and the solid contours are the 68\% confidence levels. 
We use the uniform prior to measure the upper limit for the UMO mass, of which the range is from 0 to $10^{-7}  M_{\rm \sun}$. The reason of this choice is explained in the main text.  The drop of posterior in the low-mass end is due to the choice of prior, and does not indicate a detection of non-zero mass.}
    \label{fig:posterior1}
\end{figure*}

\begin{figure}
	\includegraphics[width=\columnwidth]{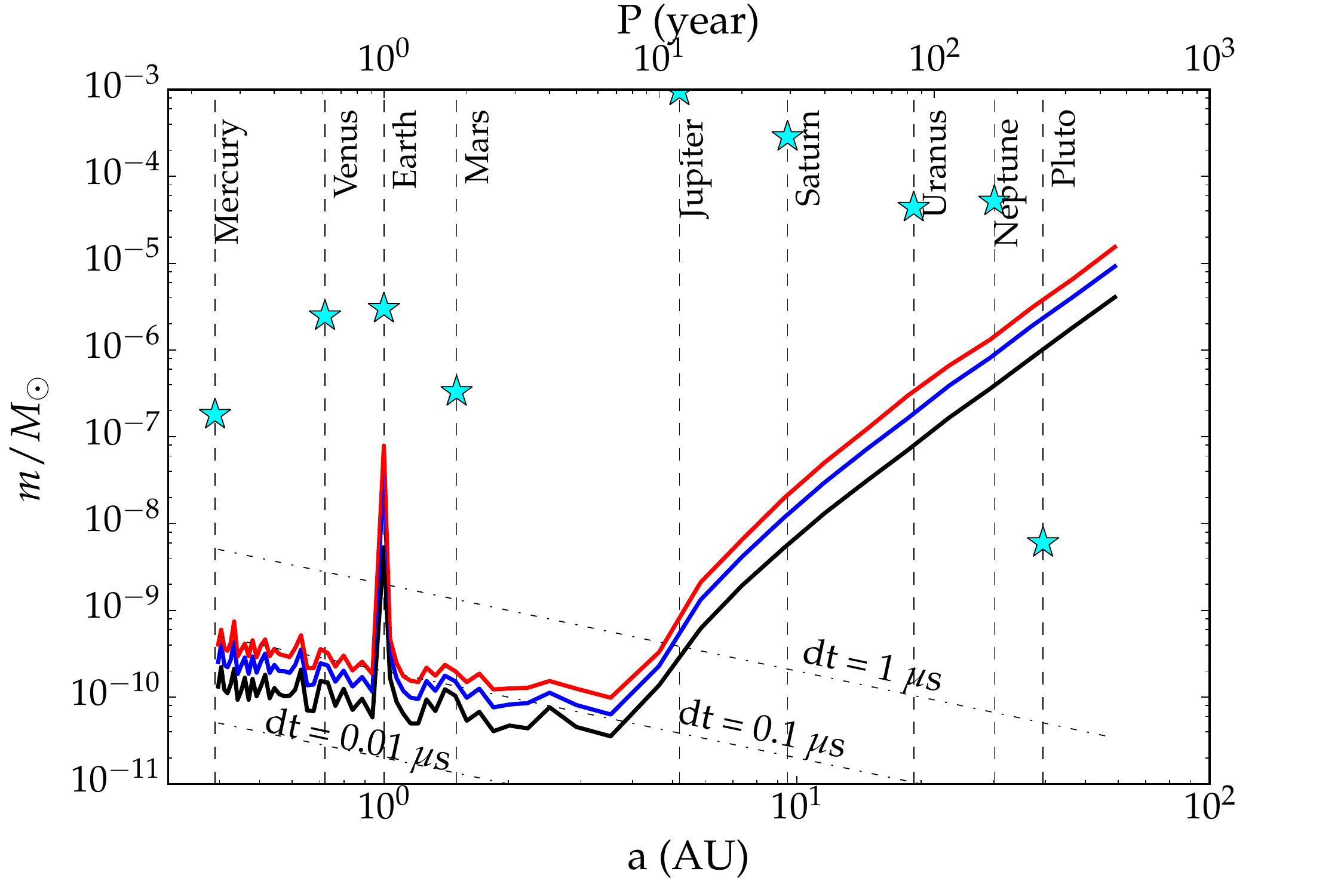}
		\caption{The upper limits for the mass of UMO at different distance to 
		the SSB are shown in solid curves, with confidence levels of 
		$1\sigma$(black), $2\sigma$(blue) and $3\sigma$(red). The vertical dashed lines 
		indicate the semi-major axis of the major planets in the Solar system, and 
		the star marks the corresponding mass. The dot-dashed lines
		show the expected amplitude ($\frac{ma}{M_{\odot}c}$) of the residuals induced by
		a mass at given semi-major axis.}
    \label{fig:upper_limit}
\end{figure}

In the second step, we focus on getting the upper limit for the mass of UMOs. 
It is more informative to know such upper limits as function of the semi-major axis 
($a$). We therefore go through a grid of semi-major-axis values and perform upper-limit
inference. For each value of $a$, we perform posterior sampling for the 
rest of the orbit and noise parameters. Using a uniform prior on the mass of UMO, we 
derive its upper limit. An example of the posterior distribution for a search for UMOs 
at $a \simeq 0.4\, {\rm AU}$ is shown in \FIG{fig:posterior1}. 
The corresponding upper limits of the UMO mass as a function of 
the semi-major axis is presented in \FIG{fig:upper_limit}. As one can see, 
with this simulated data set, any UMO with mass above $\sim 10^{-9} M_{\sun}$ should be 
excluded within $\sim 5\, {\rm AU}$ of the SSB.
The spike at $1\, {\rm AU}$ is caused by fitting for the position of the pulsar. 
This removes any sinusoidal component with an annual period, while the much smaller spike with a half-year period ($a\simeq 0.6\, {\rm AU}$) results from fitting for parallax, which removes only a sinusoidal component in phase with the Earth's orbit. 
The reduction in sensitivity for periods longer than 7 years is due to fitting for the pulse period and spin-down rate, as will be discussed in Section 4.

\begin{figure}
	\includegraphics[width=1.0\columnwidth]{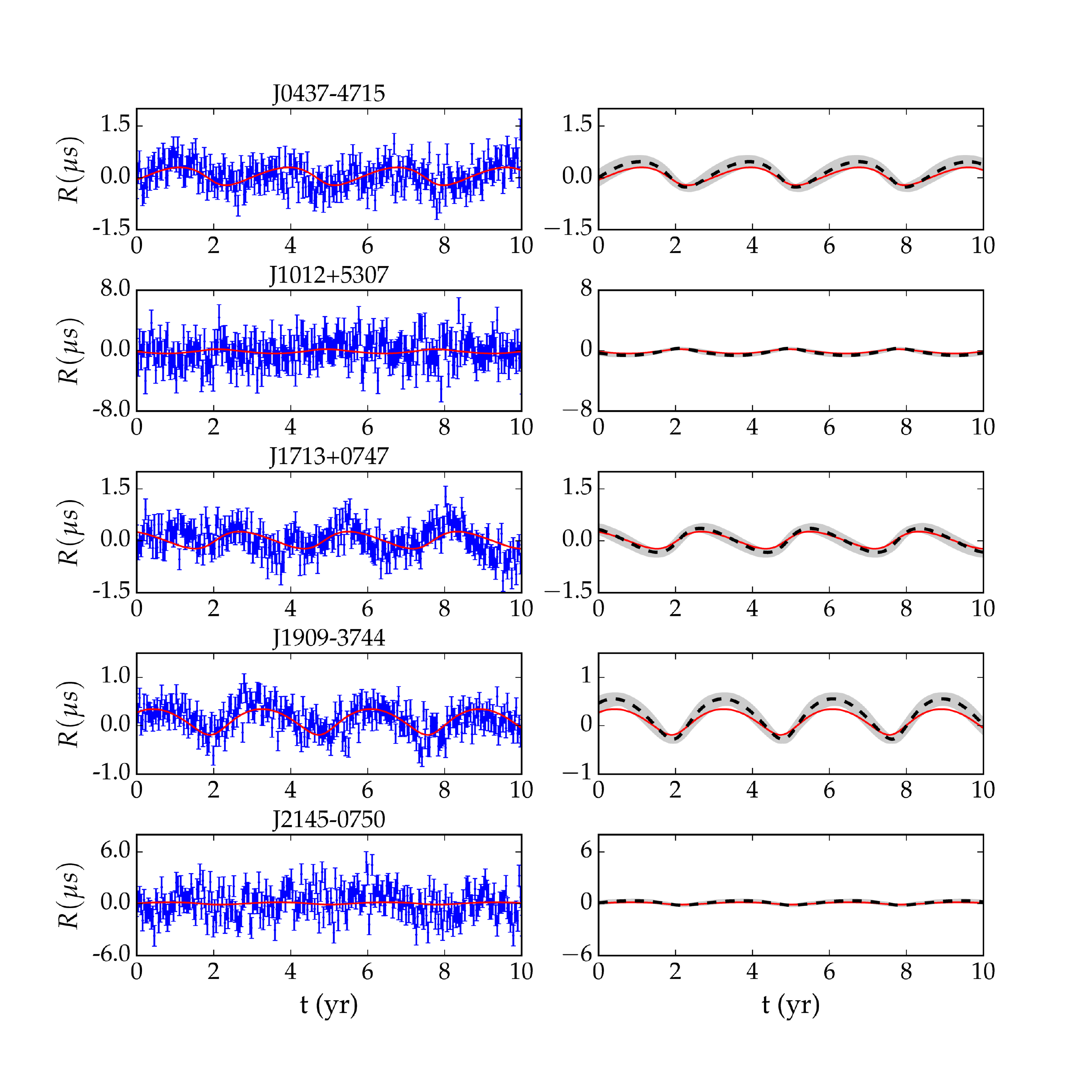}
		\caption{The simulated data, the injected and the recovered waveforms in 
		Case 2 for the five pulsars. The left column shows the timing residuals and 
		the injected waveform of UMO (the solid line with no errorbar). The 
		recovered waveforms are plotted in black dashed curves in the right column, 
		and the grey stripes indicate the 68\% confidence regions. The solid curves
		are the injections, plotted for comparison. }
    \label{fig:sim2}
\end{figure}

{\bf Data analysis and results of \textsc{case 2}}.
In this case, we inject the signal of a UMO in the simulated data set and demonstrate 
the method to measure the parameters of UMO. For the UMO signal, the semi-major 
axis is chosen to be $2\, $AU. The mass is $5\times10^{-10}\, M_{\odot}$, i.e. at the 
$3\sigma$ upper limit of \textsc{case 1}.
A moderate value of 0.3 is selected for the eccentricity.
While this value is actually larger than any other planets in the Solar system, 
we choose it to verify the ability of searching for eccentric orbit.
The remaining angle parameters are set arbitrarily to be 1 radian.
Table \ref{tab:sim2} lists those parameters. The injected UMO signal and 
simulated data are in \FIG{fig:sim2}.

\begin{figure*}
	\includegraphics[width=2\columnwidth]{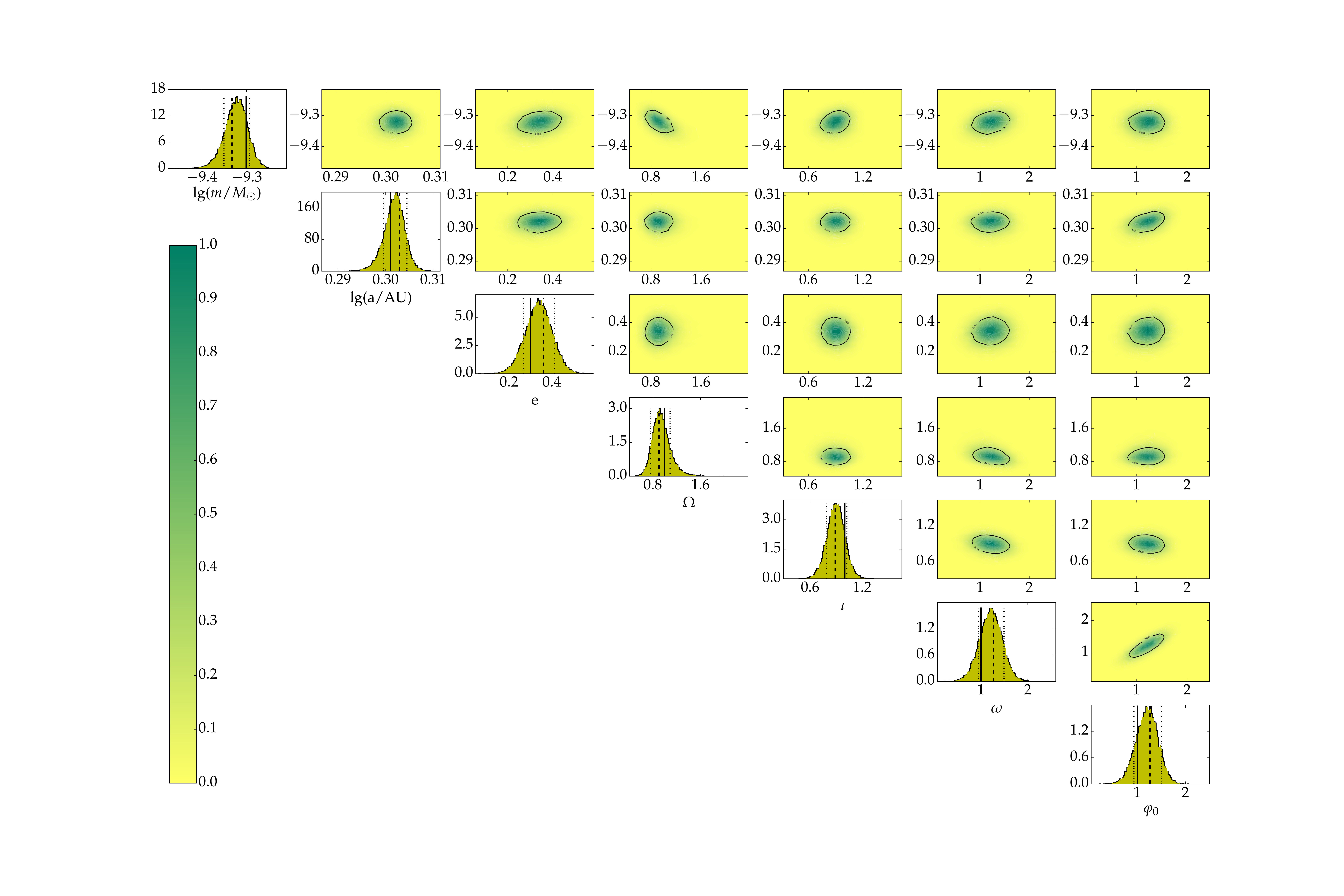}
		\caption{The marginalised posterior probability distribution of mass $m$ and 
		orbital elements $a, e, \Omega, \iota, \omega, \varphi_{0}$ in Case 2.  In 
		the diagonal 1D marginalised posteriors, the solid vertical lines indicate 
		the injected value of parameters, and the dashed lines represent the maximum likelihood estimation  
		with the 68\% confidence level marked by the dotted lines. Similar to 
		\FIG{fig:posterior1}, the upper triangular panels are the 2D marginalised 
		posterior distributions.
}
    \label{fig:posterior2}
\end{figure*}

As in \textsc{case 1}, we computed the Bayes factor 
and found that $2\log K=3.6$, 
which, again based on \citet{KR95}, is a clear preference for the 
model that includes an UMO. The posterior 
distribution for the parameter inference is shown in \FIG{fig:posterior2}.  
As a comparison, we overplot
the reconstructed waveforms using the inferred parameters
along with the injected signals in \FIG{fig:sim2}.
From these figures, one can see that for strong signals, 
the current algorithm produces compatible  
UMO parameters compared with the injection values.


\begin{table}
	\centering
	\caption{The seven parameters of UMO in Case 2. The errors here report 
	the 68\% confidence level.}
	\label{tab:sim2}
	\begin{tabular}{ccc} 
		\hline
		\hline
		Parameter& Simulation& Inference \\
		\hline
		$\log_{10}(m/M_{\odot})$ & -9.3 & $-9.33^{+0.04}_{-0.02}$ \\
		$\log_{10}(a$/AU) & 0.3         & $0.303^{+0.002}_{-0.003}$ \\
		$e$ & 0.3                       & $0.36^{+0.05}_{-0.09}$ \\
		$\Omega$ &1                     & $0.9^{+0.2}_{-0.1}$ \\
		$\iota$ & 1                     & $0.9^{+0.1}_{-0.1}$ \\
		$\omega$ & 1                    & $1.3^{+0.2}_{-0.3}$ \\
		$\varphi_{0}$ & 1               & $1.3^{+0.2}_{-0.3}$\\
		\hline
	\end{tabular}
\end{table}

\section{Analytical results and future perspectives}
\label{sec:analytical}
In this section, we derive the analytic formula for the mass upper limit of the 
UMO. We verify the analytic formula using simulations and then use the analytic 
results to investigate the capability of future PTA experiments 
in detecting the UMOs.

\subsection{Analytic formula}
The Cram{\'e}r-Rao bound is a well-studied statistical tool \citep{Fisz63} to 
determine the lowest bound on the variances of estimators. It can be regarded as 
the upper limits for the none detection, or the errorbar when the parameter is 
measured. Given the likelihood function, $\Lambda$,
the expected covariance matrix 
of parameter error is
\begin{equation}
	\left \langle \delta \epsilon_{p} \delta {\epsilon}_{q}\right\rangle  = 
	\left\langle \frac{\partial \ln \Lambda}{\partial \VEC{\epsilon}} \frac{\partial \ln 
	\Lambda}{\partial \VEC{\epsilon}}  \right\rangle_{p,q}^{-1}\,.  \end{equation}
Here, the model parameters are described 
by the vector $\VEC{\epsilon}$. They contain both the timing 
parameters $\VEC{\lambda}$ and the UMO parameters $\VEC{\xi}$.
For the Gaussian likelihood, i.e. \EQ{eq:fullik}, the above Cram\'er-Rao bound 
can be reduced to \citep{sleptian54}
\begin{equation}
	\left \langle \delta \epsilon_{p} \delta \epsilon_{q}\right\rangle 
	=\left(\frac{\partial \VEC{\Delta R}}{\partial \VEC{\epsilon}} \MX{C}^{-1} 
	\frac{\partial \VEC{\Delta R}}{\partial \VEC{\epsilon}} \right)_{p,q}^{-1}\,.
	\label{eq:sb}
\end{equation}

To proceed, we further simplify the problem by considering only white noise 
contribution and we assume that (1) the eccentricity of the UMO is small, i.e. 
$e\ll 1$, and that  (2) there are enough data points such that $N\gg1$.
The first assumption helps to get a closed form for the UMO induced signal, 
$\VEC{\Delta R}$. Under the second assumption, we can replace the summation of 
matrix indices in \EQ{eq:sb} with the continuous time integration. We then get,
\begin{equation}
\left\langle\delta \epsilon_{p} \delta \epsilon_{q}\right\rangle =   
\left(\frac{N}{\sigma^{2}T} \int_{0}^{T} \frac{\partial \Delta R}{\partial 
\VEC{\epsilon}} \frac{\partial \Delta R}{\partial \VEC{\epsilon}} \D t 
\right)_{p,q}^{-1}\,.  \label{eq:mass_error}
\end{equation}

\begin{figure}
	\includegraphics[width=3.5in]{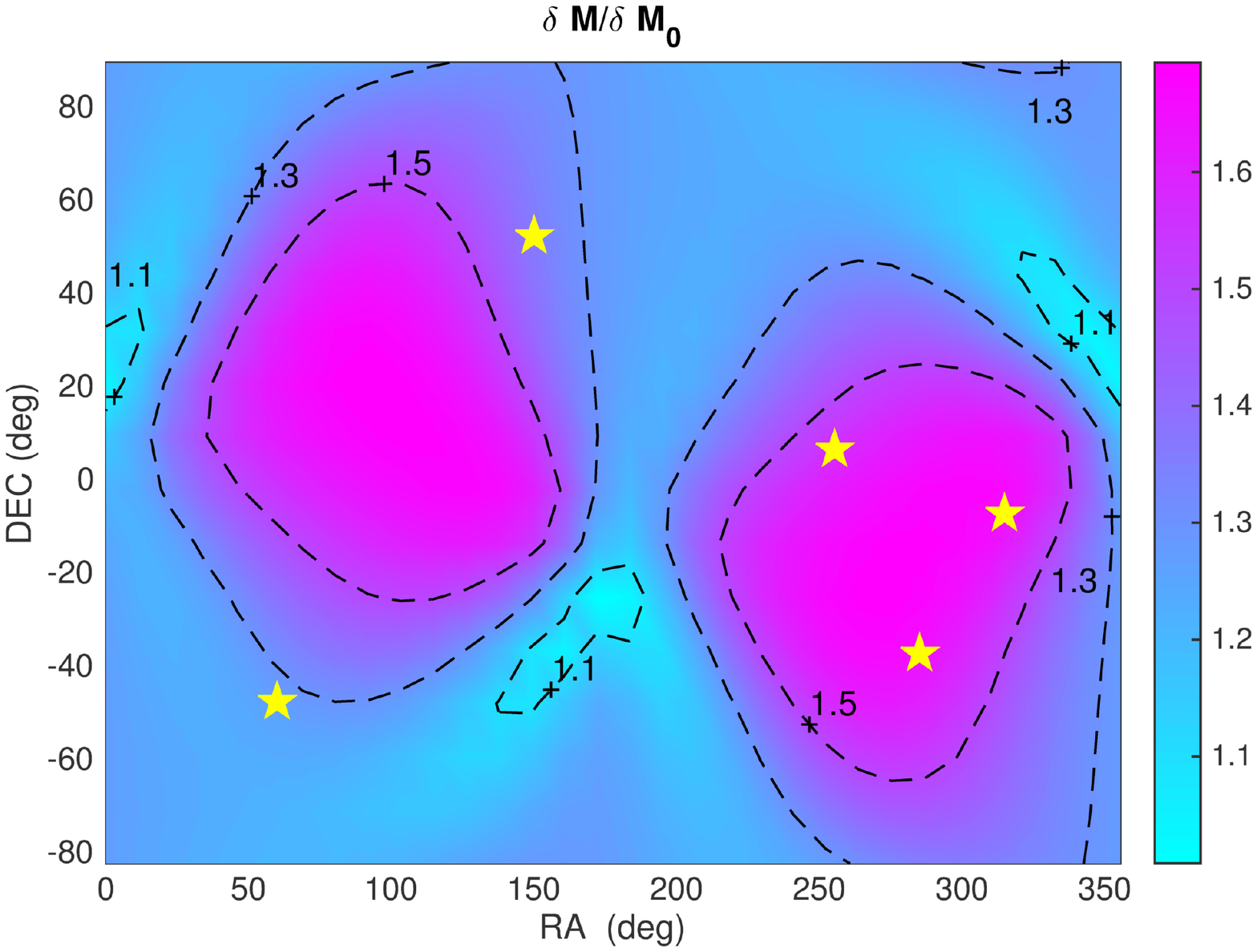}
	\caption{The geometric correction factor $\kappa$ for the five pulsars we 
	used. The yellow pentagons are the five pulsars. The colour scale together with the 
	contours indicates the value of $\kappa$. Here we plot $\kappa$ as a function 
	of right ascension (RA) and declination (Dec.) of the orbit pole, i.e. the 
	projection direction of the orbital angular momentum on the celestial sphere.  
	We see that $\kappa$ becomes larger, when the orbital pole is located around clusters of 
	pulsars. This means that the sensitivity to UMOs drops, when the 
	orbital plane is perpendicular to the pulsar direction.}
		\label{fig:geo}
\end{figure}

For the mass of UMO, \EQ{eq:mass_error} 
leads to an analytical expression at the two 
following limits of the semi-major axis:
\begin{equation}
\frac{\delta m}{M_{\odot}}=
\begin{cases}
\kappa \frac{\sqrt{2}\sigma c}{\sqrt{N}} a^{-1}, &\text{if } a<\frac{(\sqrt{GM}T)^{2/3}} 
{(120\sqrt{7})^{2/9}}; \\
\kappa \frac{120\sqrt{14}\sigma c}{\sqrt{N}(\sqrt{GM}T)^{3}} a^{7/2}, &\text{if } 
a>\frac{(\sqrt{GM}T)^{2/3}} {(120\sqrt{7})^{2/9}}.
\end{cases}
\label{eq:analexp}
\end{equation}
where $\sigma$ is the effective root-mean-square level of pulsar noise, 
\begin{equation}
        \sigma = \left(\sum_{i=1}^{N_{\rm psr}} \sigma_i^{-2} \right)^{-0.5}\,.
	\label{eq:effsig}
\end{equation}
In the above equations, $N$ is the average number of observation epochs per pulsar, $T$ is the average
length of observation, and $\kappa$ is the sky sensitivity. The latter is the geometric 
correction factor for taking into account the projection 
of the signal to the pulsar direction, as in \EQ{eq:delay}. 
The sky sensitivity, $\kappa$, 
approaches unity, when the number of pulsars, $N_{\rm psr}\gg 1$. For limited number 
of pulsars, $\kappa$ can be treated as a factor of 2, as we do
for the case of five pulsars in our analysis.
The sensibility of this approximation can be illustrated by plotting the value of 
$\kappa$ for the five pulsars we used as shown in \FIG{fig:geo}. 

In \EQ{eq:analexp}, there are two cases depending on the value of $a$. For the 
case of a small $a$, the frequency of the UMO signal is high such that the UMO 
signal and the quadratic pulsar spin-down signal are uncorrelated. The error of 
UMO mass is inversely proportional to $a$ because of \EQ{eq:ds}, and 
in this regime, the UMO 
located farther to the SSB is easier to detect. In the second regime, when $a$ 
becomes larger, the data span is not long enough to cover several orbital 
periods, so the short-duration sinusoidal function is correlated with the 
quadratic pulsar spin-down signal. The UMO signal is no longer 
periodic in the data. The $a^{7/2}$ dependence comes from the cubic function left in the 
residuals due to the fitting of pulsar period and period derivative. 

\begin{figure}
	\includegraphics[width=\columnwidth]{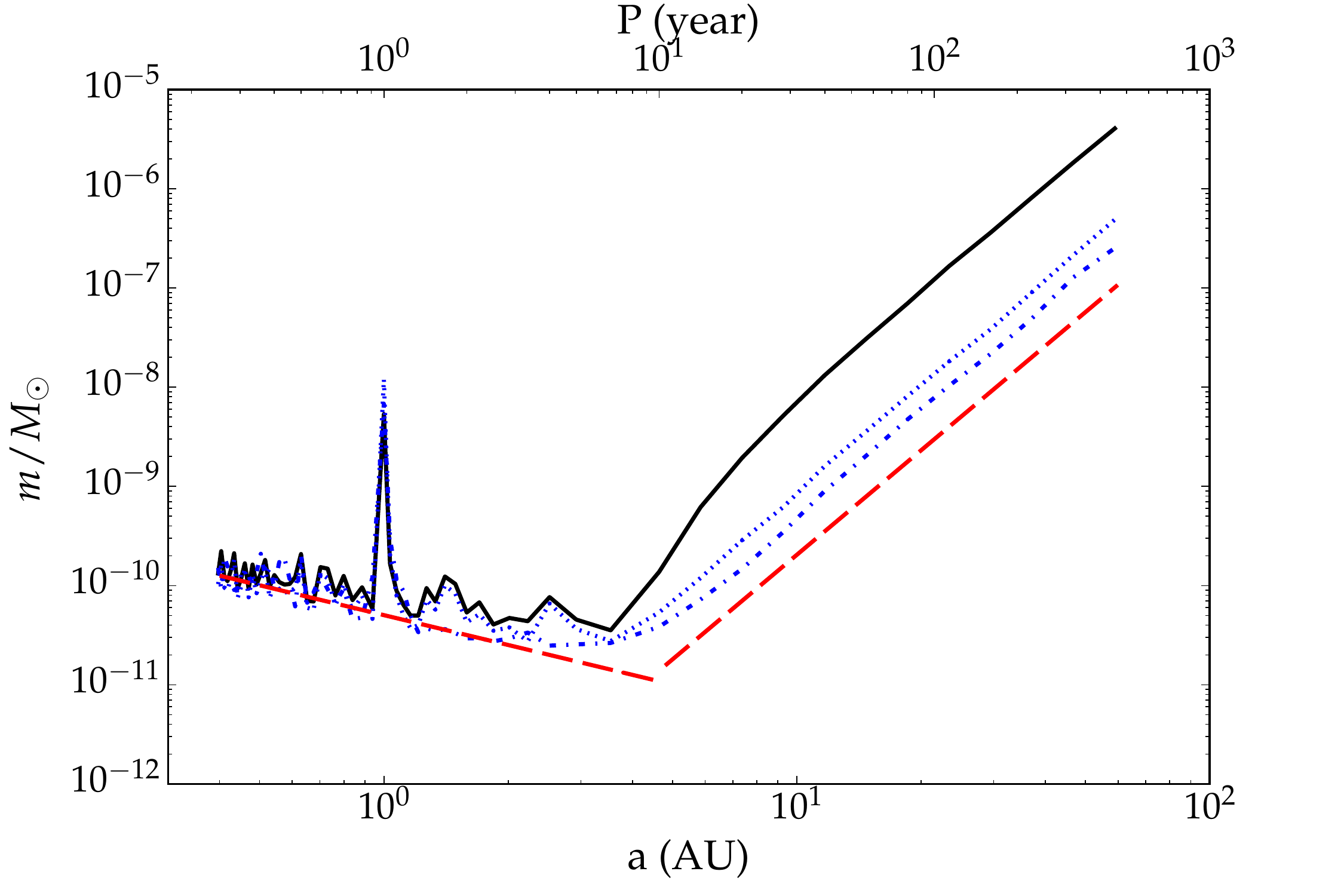}
		\caption{The analytical upper limits for the mass of UMO. The red dashed 
		lines show the approximations in the limits of $T\gg a^{1.5}$ and $T\ll a^{1.5}$ as in 
		\EQ{eq:analexp}. The 68\% upper limit in Fig.~\ref{fig:upper_limit} is
		plotted in a black solid curve for comparison. The blue dotted line is the 
		upper limit when the eccentricity prior is confined to $0\le e<0.1$. The 
		blue dot-dashed line is the upper limit when we do not fit for red noise 
		parameters.}
    \label{fig:estimation}
\end{figure}

To assess the predictive power of \EQ{eq:analexp},
we employ it to calculate analytically
the UMO-mass sensitivity curves for the \textsc{case 1} simulations
and compare it with the results from the Bayesian analysis.
The results are shown in \FIG{fig:estimation}. One can 
see that such an analytical expression, although much simplified, still encapsulates 
the major features of the sensitivity curve. The deviations between the analytical 
expression and the numerical results become significant, when $a$ is large. In this 
regime, the UMO orbital period is larger than the data length, and the estimations of mass upper 
limit become highly affected by the choice of prior and red noise modelling. The 
analytical expression gives the correct power index, i.e. $\delta m\propto 
a^{7/2}$, but the numerical factor becomes less reliable.

\begin{figure}
	\includegraphics[width=1.1\columnwidth]{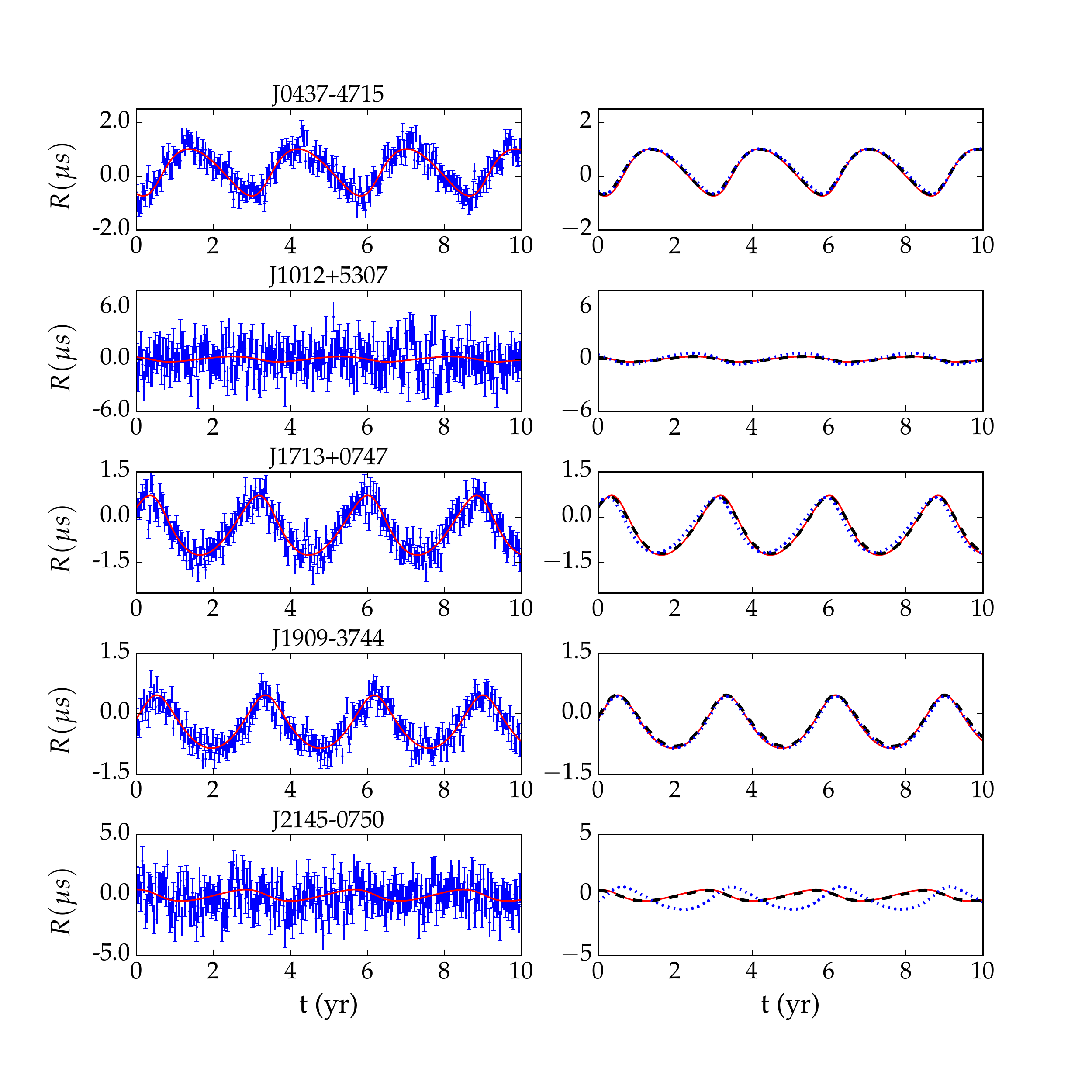}
    \caption{The simulated data and recovered waveform of an
		ill-conditioned example. The left column shows the timing
		residuals and the simulated waveform of UMO (red solid line). The
		waveform recovered from the two different sets of parameters are plotted
		in dashed black and dotted blue lines in the right column. As one can see, 
		when the number of pulsars with good signal-to-noise ratio is limited, 
		multiple solutions will be allowed. } 
		\label{fig:sim3}
\end{figure}

\begin{figure*}
	\includegraphics[width=2\columnwidth]{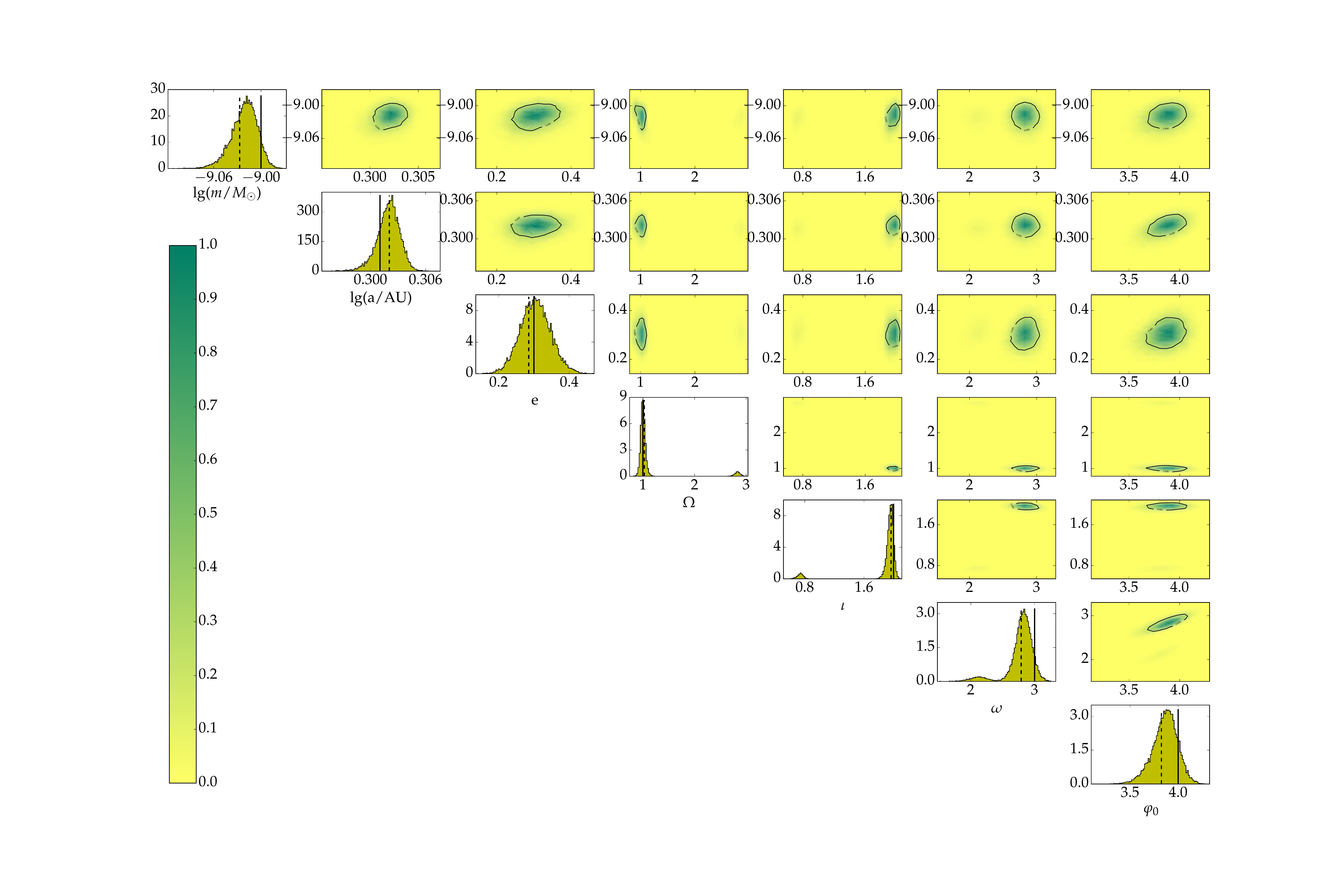}
	\caption{The same as Figs \ref{fig:posterior1} and \ref{fig:posterior2}, but for an 
	ill-conditioned simulation. The two peaks in the posterior of $\Omega, \iota$ 
	and $\omega$ indicate the multiple solutions.}
    \label{fig:posterior3}
\end{figure*}

The Cram\'er-Rao bound will fail for the situation, when a unique un-biased 
estimator is not available \citep{LWK11}. For the current UMO problem, this 
happens, when the number of pulsars is limited. In certain cases, a limited 
number of pulsars have much better precision than the rest of pulsars in the timing 
array. This effectively reduces the number of pulsars contributing to the 
analysis. We show an ill-conditioned 
example in \FIG{fig:sim3}, and the corresponding posterior distributions
from the Bayesian analysis in \FIG{fig:posterior3}. Although the 
recovered waveform is very different for PSR J2145$-$0750, the timing precision 
limits us from differentiating the two sets of parameters. In general, three parameters 
can be measured from the signal of one pulsar, namely the amplitude, frequency 
and phase. We will need more than three pulsars to measure the seven parameters 
describing the UMO.

\subsection{Prediction}
In the near future, discoveries of more stable millisecond pulsars and new 
commissions of advanced instruments will continuously increase the quality of 
pulsar-timing data. The \emph{Five-hundred-meter Aperture Spherical radio
Telescope} \citep[FAST;][]{Nan11}, the \emph{QiTai 110m radio Telescope}  
\citep[QTT;][]{Wang17} and the \emph{Square Kilometre
Array} \citep[SKA;][]{KS10}, will significantly improve the timing precision for 
a large number of pulsars. We expect the upper limit for the UMO mass will be more 
stringent.

\begin{figure}
	\includegraphics[width=\columnwidth]{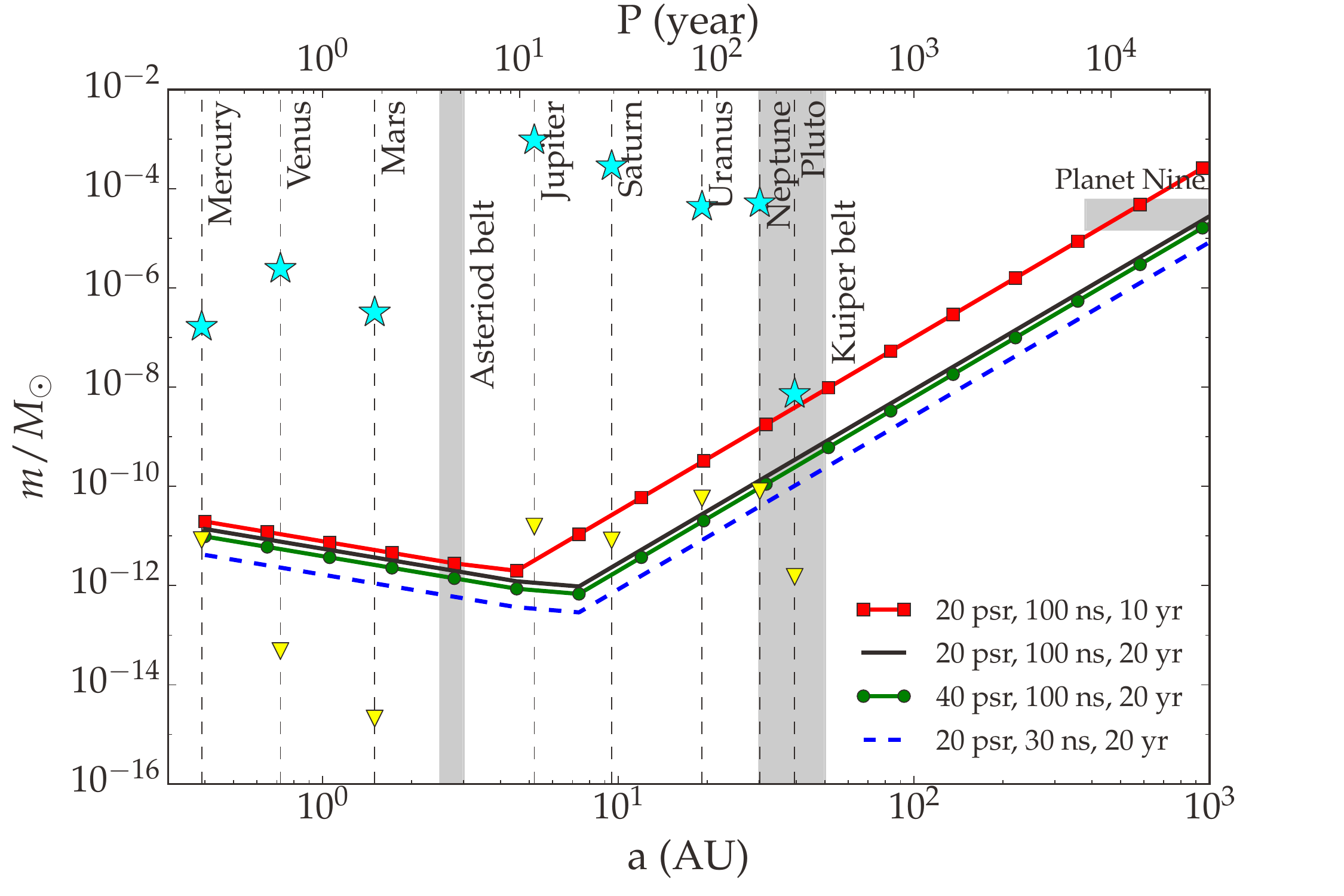}
		\caption{The predicted upper limits for the mass of UMO in four cases, with 
		different combinations of the number of pulsars, timing precision and data 
		span. For all cases, the cadence of observation is chosen to be two weeks. 
		The vertical dashed lines indicate the semi-major axis of the major planets in the Solar system, and  
		the star marks the corresponding mass. 
		The official IAU uncertainties on the planetary masses are plotted as triangles for comparison \citep{IAU09}.
		The mass at 1$\, $AU will be practically unconstrained because of fitting pulsar position, so the parameters of Earth are not plotted.
		The grey regions show the positions of the asteroid belt, Kuiper belt and the parameter space of Planet Nine \citep{BB16b}.
		Note that the prediction is for circular orbit, and the upper limit will be larger for an eccentric orbit. }
    \label{fig:prediction}
\end{figure}

We estimate the future upper limits of the UMO mass using \EQ{eq:analexp}, and 
the results are summarized in \FIG{fig:prediction}. 
One can see that with 10-year pulsar timing for 20 pulsars to the precision of 100 ns, 
we can push the mass upper limits to $10^{-11}$ to $10^{-12} M_{\rm \sun}$, i.e.  
$10^{-8}$ to $10^{-9}$ Jupiter mass. Other cases show the improvement of upper limits
with longer data span, more pulsars and increased precision. The upper limits can be an order 
of magnitude better, if we use 20-year data of 20 pulsars with a precision of 30 ns. Since the timing precision of 
30$\, $ns over 20 years is unlikely due to noise processes in pulsars, the result will 
probably look that good only for UMOs with periods of $\sim5$ years. 
The upper limit is also the precision of measurements, so we expect that future PTAs 
will measure the mass of Jovian system with a fractional precision of $10^{-8}$ to $10^{-9}$, which is 
comparable to the existing IAU uncertainties~\citep{IAU09}. 

\section{Discussions and Conclusions}
\label{sec:conclusion}

We have developed a method to search for UMOs in the \SoS using PTA data. 
Our algorithm is based on the Bayesian data-analysis framework, 
where the detection significance is evaluated using Bayes factor and the 
parameter inference is performed using posterior sampling. We have verified the 
method using simulated data sets. The current method is capable of producing the 
upper limit for the mass of UMOs and to measure the orbit elements.
As we have demonstrated, the parameter inference matches the injection values in 
the simulation. We have also derived the analytical expression for the upper 
limits of UMO mass. The upper limits using the Bayesian inference agree well 
with our analytical expression. With this method, we have estimated the future 
perspectives of detecting the UMOs using pulsar timing array. The method has 
different selection effects compared to the traditional planet detections, e.g.  
optical surveys, that even the invisible exotic objects can be detected, as long 
as they are massive.

\citet{Champion10} measured the mass error
of the known planets, based on the \SoS ephemeris DE421. By employing a 
dynamical model for the orbit, our method measures the properties of 
UMOs, both the mass and
orbital elements. The Bayesian framework helps to simultaneously analyze the 
signal of UMOs, pulsar timing model in the presence of other noise processes. 
We have assumed Keplerian orbits for the UMOs and neglected all perturbations.
Here, the assumption saves us from implementing the full dynamic
modelling of \SoS as in \citet{Almanac10}. 
In our model, we neglect all propagation effects of pulsar signal due to the UMO other than R\o mer delay. 
The higher order terms, e.g. Shapiro delay and gravitational red shift of UMO, will be not be measurable for objects much lighter than the Jupiter. 
While obviously having the advantage of fully exploring the 
orbital parameters, the
method is currently practically limited to study light objects not in orbit
with other planets. Nevertheless,
the algorithm presented can serve as a basis on which
we can found attempts to perform full dynamical modelling in the future. 
It is noteworthy that inaccuracies in the used \SoS ephemeris,
are identified as one of the main sources of correlated noise
in PTA data that impede efforts for direct nHz GW detections \citep[e.g.][]{thk+16, Taylor17, WHC17}.
Modelling approaches such 
as the one presented in this work, can help in efforts to
mitigate this noise and improve the PTA sensitivity to GWs \citep{LTM15, LGW16}.
We also note, that while discoveries of previously 
unknown bodies in the solar system
with PTA blind searches may be difficult,
the use of more evolved dynamical models 
may allow in the future PTAs to contribute in imposing meaningful 
constraints on the parameter space of independently
proposed unknown planets \citep[e.g.][]{bb16, BB16b}, as shown in \FIG{fig:prediction}.

Both the pulse period $P$ and period derivative $\dot{P}$ are fitted in the timing model, which absorbs the linear and quadratic signals. 
In this way, our method is not 
sensitive to the acceleration of the Solar system, and it searches for the `jerk' in the 
timing signal for long-period planets, i.e. we search for the time 
\emph{derivative} of \SoS acceleration for the second case in \EQ{eq:analexp}.  
There are works to constrain the \SoS acceleration directly.
\citet{ZT05} proposed to detect the \SoS acceleration using $\dot{P}$ distribution of millisecond pulsars (under 
the assumption of position-independent distribution) or orbital period derivative of binary pulsars.
\citet{VB08} and \citet{DV08} timed the binary PSR J0437$-$4715 and determined 
the upper limit of \SoS acceleration 
using orbital period derivative.  

Our analytical expression for the mass upper limit is derived using Cram\'er-Rao 
bound. Since it theoretically predicts the best possible upper limit for
any unbiased estimator, it is a very useful tool to cross check the data-analysis
as well as to make predictions to help planning future observations. 
As we see, timing 20 
pulsars to the precision of 100 ns, will rule out any unknown objects with mass 
of $10^{-11}$ to $10^{-12} M_{\rm \sun}$ within 10 AU around the Sun. For dark 
matter clumps, this will be a factor of $10^{2}$ better than the current limit 
\citep{PP13, PP13b}. The PTAs become sensitive tools to study the \SoS mass distribution and 
dynamics. We expect that advanced instruments (e.g. FAST, SKA, and QTT) in the 
future will benefit the field.

\section*{Acknowledgments}
This work was supported by XDB23010200, National Basic Research Program
of China, 973 Program, 2015CB857101 and NSFC U15311243, 11690024, 11373011. We are also supported by the MPG funding for the Max-Planck Partner Group. The
computation was performed using the cluster \textsc{Dirac} in KIAA and
the \textsc{Tianhe II} supercomputer at Guangzhou supported by Special Program for Applied Research on Super Computation of the NSFC-Guangdong Joint Fund (the second phase) under Grant No.U1501501. We thank Joris Verbiest, William Coles and Stephen Taylor for  helpful comments.




\newpage
\bibliographystyle{mn2e}
\bibliography{ms} 







\bsp	
\label{lastpage}
\end{document}